\documentclass{llncs}
\usepackage{alltt}
\usepackage{graphicx}
\usepackage{wrapfig}
\usepackage{lettrine} 
\usepackage{amsmath}  
\usepackage{amssymb}  
\usepackage{amssymb}  
\usepackage{textcomp} 
\usepackage{xcolor}   

\newenvironment{code}{\begin{small}\begin{alltt}}{\end{alltt}\end{small}}

\makeatletter
\@ifpackageloaded{txfonts}\@tempswafalse\@tempswatrue
\if@tempswa
  \DeclareFontFamily{U}{txsymbols}{}
  \DeclareFontFamily{U}{txAMSb}{}
  \DeclareSymbolFont{txsymbols}{OMS}{txsy}{m}{n}
  \SetSymbolFont{txsymbols}{bold}{OMS}{txsy}{bx}{n}
  \DeclareFontSubstitution{OMS}{txsy}{m}{n}
  \DeclareSymbolFont{txAMSb}{U}{txsyb}{m}{n}
  \SetSymbolFont{txAMSb}{bold}{U}{txsyb}{bx}{n}
  \DeclareFontSubstitution{U}{txsyb}{m}{n}
  \DeclareMathSymbol{\aleph}{\mathord}{txsymbols}{64}
  \DeclareMathSymbol{\beth}{\mathord}{txAMSb}{105}
  \DeclareMathSymbol{\gimel}{\mathord}{txAMSb}{106}
  \DeclareMathSymbol{\daleth}{\mathord}{txAMSb}{107}
\fi
\makeatother

\def\QED{\raisebox{0.8ex}{\framebox{\kern0.2ex}}}

\title{Safe Compilation for Hidden Deterministic Hardware Aliasing
and Encrypted Computing}
\author{
Peter T. Breuer\inst{1}
}

\institute{
Hecusys LLC, Atlanta, GA, USA. Email: ptb@hecusys.com
}

\begin{document}
\raggedbottom

\maketitle

\vspace{0.5in}

\begin{abstract}
Hardware aliasing occurs when the same logical address sporadically
accesses different physical memory locations and is a problem
encountered by systems programmers (the opposite, software aliasing,
when different addresses access the same location, is more familiar to
application programmers).  This paper shows how to compile so code works
in the presence of {\em hidden deterministic} hardware aliasing.  That
means that a copy of an address always accesses the same location,
and recalculating it exactly the same way also always gives the same
access, but otherwise access appears arbitrary and unpredictable.
The technique is extended to cover the emerging technology of
encrypted computing too.
\end{abstract}

\section{Introduction}
\label{s:Intro}

{\em Hardware aliasing}   describes ``the
situation where, due to either a hardware design choice or a hardware
failure, one or more of the available address bits is not used in the
memory selection process.''\,\cite{Barr98} Its effect 
used to be familiar to programmers and users alike, as the `DLL hell'
that the old 16-bit versions of Windows were prone to.  Dynamic linked
libraries (DLLs) were problematic for many reasons, but one was that
different versions of the same library loaded at the same memory address
and all applications referenced the same in-memory copy.  So if one
application loaded one version of the library then another 
loaded another version, the first application would unpredictably, as
far as the program and user were concerned, find itself in the second
library's code.  A virtual extra address bit was the second application
loading its version of the library.

DOS users were more familiar with the situation, as
expanded memory managers (for memory beyond 1MB) such as QuarterDeck's
QEMM \cite{Glosserman85} remapped memory so the video graphics and
bootstrap (BIOS) code both shared addresses with random access memory (RAM).
What a program accessed at runtime at a given address depended on the
memory manager, and that decided heuristically.  In consequence, program
code needed to use standard memory access sequences in order to
trigger expanded memory reliably.

The recognisable symptom of hardware aliasing is that what looks
to programs like the
same memory address (the {\em logical} address), sporadically accesses
different memory locations (the {\em physical} address).  With
memory-mapped input-output (IO), that may mean different peripheral
devices.

Modern
applications programmers are more familiar with the opposite
{\em software aliasing}, in which the same physical
resource -- memory location or peripheral device -- is accessible via
different logical addresses. But hardware aliasing has not gone away so much
as become irrelevant as hardware has evolved
to present a more unified view to software.  The paradigm is
{\em oblivious} RAM (ORAM)
\cite{ostrovsky1990,ostrovsky1996,ostrovsky2013}, a secure RAM solution
since the 90s in which memory contents are internally moved about and
locations rewritten sporadically and independently to frustrate cold boot
attacks \cite{Gruhn2013} (freezing the memory sticks to make
the temporary arrangement of electrical charge in RAM last long enough to
be analysed when powered off).  That internal
re-aliasing is not externally visible.

Hardware aliasing arises nowadays
notably in the context of {\em encrypted computing}
\cite{fletcher2012,BB13a,oic,heroic,BB16a,BB16b,DBLP:journals/corr/abs-1811-12365}.  That emerging
technology is based on a processor in which inputs, outputs, and all
intermediate values exist only in encrypted form.
With an appropriate machine code instruction set and
compiler, {\em cryptographic semantic security} \cite{Goldwasser1982}
obtains for runtime user data in that environment,
relative to the security of the
encryption \cite{BB18c} (that means no deterministic or stochastic
attack has any more likelihood than chance of successfully
guessing any bit of user data, accepting the
supposition that the encryption embedded in the machine is independently
secure).  The lemma there is ``encrypted computing does not compromise
encryption." The technology has potential for widespread adoption as
near conventional speeds are being attained in recent prototype
designs \cite{cryptoblaze18,BB18b}.
Because encryption is one-to-many, many bitwise different encryptions
of the same memory address are
presented to the memory device by the processor (memory
is not privy to the encryption).  From the point of view
of the program, hardware aliasing occurs.  The programmer
says to write $x$ at address $123$ and the encrypted value 0x123456789a of
$123$ is passed to the memory management unit (MMU) at runtime.  If the
programmer later says to read via address $123$, a different value
0xa987654321 that is an alternative encryption of $123$ may be passed as
an address to the MMU, and some content $y$ different from $x$ returned.

An applications programmer must be able to remain ignorant of the
hazard.  It turns out that there is a compiler mechanism that admits
that, and this paper sets it out.  The present authors have made 
suggestions in the past as to what to do \cite{BB14c}, but experience 
has shown those to be too delicate in practice (array accesses
via pointer vs.\ index become incompatible).  The technology developed in
this paper is more computationally inefficient but requires no semantic
changes to source code, and has been deployed and tested (c.f.  the
HAVOC compiler suite for {\sc ansi} C \cite{ansi99} and encrypted
computing at \url{http://sf.net/p/obfusc}).

The solution arrived at depends on the underlying determinisn in
processors.  Processors are designed to produce repeatable results.
Thus the one-to-many-ness of the encryption in the encrypted programming
context is a function of hidden extra `padding' and/or `check' bits that
accompany the data beneath the encryption.  Those bits are calculated
deterministically by the processor, and depend on the data and the
sequence of operations it goes through.  The outcome is so-called {\em
hidden deterministic} hardware aliasing.  The features of that are:
\begin{enumerate}
\item
A machine code copy instruction copies the physical bit sequence
exactly, such that a copied address accesses the same memory location as
the original;

\item
repeating the same sequence of operations produces an address that has
exactly the same bit sequence and accesses the same location;

\item
logically different addresses always have different physical bit sequences.
\end{enumerate}
Condition~1 (`faithful copy') means the compiler can save an address for
later use after writing through it, and it will work to retrieve the
written value.  The address must not be altered, not even by adding
zero, as calculation alters the hidden padding or check bits and hence
the physical representation as a sequence of bits, which then fails to
access the same memory location as before.

Condition~2 (`repeatability') allows for some calculations on
addresses, so long as they are repeated exactly each time.  That is useful
because the machine code instruction to read or write a memory location
takes a base address $a$ in a register and adds a displacement $d$
embedded in the instruction to get address $a+d$ for the access.  It is
impossible to avoid the processor doing at least that one addition.  But
that does not matter because the same calculation is repeated each time
the address is needed, with the same sequence of bits resulting.

Condition~3 (`no confusion') guarantees that the
representations as physical sequences of bits of what look like different
addresses to the processor and program do not step on each other.
Equality as measured programmatically partitions the space of physical
bit sequences representing addresses into disjoint equivalence classes.
Each is the set of possible representations as different bit
sequences for a single logical address.  That abstraction does 
capture the situation where extra check bits or padding are ignored by
the processor's arithmetic.


The layout of this paper is as follows.  Section~\ref{s:HA} explains how
to compile to obtain code `safe against hardware-aliasing' in the most
general context.  Section~\ref{s:EC} details the extension for
compilation in the encrypted computing context.  The compiler must
vary the generated object code maximally while maintaining the same
structure (and the same runtime trace structure).  The technique
introduced in Section~\ref{s:HA} copes with apparently unreliable
{\em addressing} that is fundamentally deterministically generated and the
technique of Section~\ref{s:EC} extends that to both generate and cope
with apparently unreliable {\em data}.

\section{Compilation for Hardware Aliasing}
\label{s:HA}

The compiler principle followed in this paper is that each address that
is written is intended to be saved for later read, as per Condition~1
of Section~\ref{s:Intro}.  The conundrum in that is that it is saved at
an address, which must also be saved, and so on recursively.
Condition~2 puts a backstop on the potentially infinite recursion,
allowing addresses that are calculated at compile time by the compiler
to be used instead.  But a finite set of addresses cannot suffice for nested
function calls to unbounded depth, so the runtime stack must be
involved.  The real first problem to be solved is how to manipulate the
{\em stack pointer} so addresses and other data might be saved and
recovered reliably from stack.

\subsection{Stack Pointer 101}

Standard compiler-generated code decrements the stack pointer
register {\bf sp} by the amount that will be
needed for local storage in the function immediately on entry to the
function body, and increments it again just before exit:
\begin{center}
\begin{minipage}{0.5\textwidth}
\begin{code}
\rm call to{\em function}
 \dots
{\em function} code start:
{\color{red}\rm decrement}{\bf sp}
 \dots
{\color{red}\rm increment}{\bf sp}
\rm return
\end{code}
\end{minipage}
\end{center}
That does not work in a hardware aliasing environment, because the
increment does not restore exactly the physical representation
originally in the stack pointer register. Instead,
the caller gets back a different set of bits that means the same thing
to the processor. Being different, it references a different area via the MMU.

The {\em frame pointer} register {\bf fp} must be co-opted to
save the stack pointer in, and the stack pointer restored from it on
exit:
\begin{center}
\begin{minipage}{0.5\textwidth}
\begin{code}
\rm{\it function} code start:
{\color{red}\rm copy}{\bf sp} to{\bf fp}
\rm decrement{\bf sp}
 \dots
{\color{red}\rm copy}{\bf fp} to{\bf sp}
\rm return
\end{code}
\end{minipage}
\end{center}
That is the typical form of unoptimised function code from a compiler.
Compiler optimisation ordinarily replaces it with the previous (stack
pointer -only) code.  The GNU {\em gcc} compiler (for example) with
{\bf-fno-omit-frame-pointer} on the command line turns off optimisation
and produces the code immediately above.

That is not quite perfect because the caller's frame pointer register must
also be saved and later restored, as follows:
\begin{center}
\begin{minipage}{0.5\textwidth}
\begin{code}
\rm{\it function} code start:
{\color{red}\rm save} old{\bf fp} to 1 below{\bf sp}
\rm copy{\bf sp} to{\bf fp}
\rm decrement{\bf sp}
 \dots
\rm copy{\bf fp} to{\bf sp}
{\color{red}\rm restore} old{\bf fp} from 1 below{\bf sp}
\rm return
\end{code}
\end{minipage}
\end{center}
Saving below the caller's stack pointer puts it in the callee's stack area
(`frame'), so its stack requirement (`frame size') is over-stated by
one in the decrement to make room for it.  As many as the compiler wants
of the caller's registers can be saved at the top of the callee's 
frame. The application binary interface (ABI) document for the platform
specifies which registers the callee must save, and which the caller code must
expect may be trampled and must save for itself.

Rather than decrement the stack pointer again for blocks of code
with their own local declarations within a function declaration, the
compiler holds the stack pointer constant and reserves space for 
subframes within the function frame.  That makes the function frame
uncomplicated to access from within sub-blocks.

The final code above works with hardware-aliasing.  It sets up storage
for the function on the stack (the `local frame') that can be reliably
addressed as ${\it sp}+d$ from within the function code in a
hardware-aliasing environment, where $d$ is a displacement between 0 and
the frame size, provided as a constant in a {\em load} or {\em store}
machine code instruction, and {\em sp} is the stack pointer register
value.

\subsection{Accessing Variables}
\label{ss:Var}

Given the setup described above for the stack pointer, accessing
function local variables is simple.  A word-sized local variable {\bf x}
is assigned a position $n$ on the stack and the compiler issues a
load instruction to read from there to register\,$r$:
\begin{center}
\begin{minipage}{0.5\textwidth}
\begin{code}
lw{\it r}{\it n}({\bf{sp}})  \#\rm load from offset \(n\) from{\em sp}
\end{code}
\end{minipage}
\end{center}
The processor does addition ${\it sp}+d$ in executing the instruction,
but repeats that calculation at every access, so by Condition~2
of Section~\ref{s:Intro} the same sequence of bits for 
the address is produced every time, and it accesses the same spot in
memory via the MMU.  To write the variable, a {\em store} instruction
is used instead:
\begin{center}
\begin{minipage}{0.5\textwidth}
\begin{code}
{\color{red}sw}{\it n}({\bf{sp}}){\it r}  \#\rm store to offset \(n\) from{\em sp}
\end{code}
\end{minipage}
\end{center}
For global variables, which reside at a compiler pre-decided address $a$
in (heap) memory, the compiler offsets from the zero register {\bf zer}
instead of {\bf sp}:
\begin{center}
\begin{minipage}{0.5\textwidth}
\begin{code}
lw{\it r}{\it a}({\color{red}\bf{zer}}) \,\#\rm load from address \(a\)
\end{code}
\end{minipage}
\end{center}
The zero register contains a fixed zero value (that speeds up
computations as zero is frequently needed). The 
address presented to the MMU by this instruction is $a+0$, which is a
different sequence of bits to $a$ (representing the same value
to the processor), but it is what is always presented
so the same memory location is always accessed.
The compiler also pre-calculates the bit sequence for $a+0$ and puts
data into the executable/loadable file (ELF) for the program loader to
load at the location before program start, to provide an initial value.

Variables in the parent's frame may also be accessed.  If the function
is defined within another function, it is an {\em interior} function,
and the parent's local variables should be notionally in scope for it.
The parent's frame pointer is handed down at runtime in the {\bf c9}
register (that register is specific to the platform ABI).  The register
is saved through successive daughter function calls along with the frame
pointer register.  Then a load or store instruction using {\bf c9}
instead of {\bf sp} or {\bf zer} reliably accesses the parent function's
local variables.

\subsection{Accessing Arrays}

The elements of array $\bf a$ can in principle be accessed either via
a load or store instruction with fixed displacement
$n$ from the array address $a$ (that is `$a[n]$'),
or via a pointer with value $p$ that ranges through
the array starting at $a$ and steps through the elements until
the desired one is reached, at which point a load or store instruction
with displacement 0 from the pointer (`$p[0]$') is applied \cite{BB14c}.
The two calculations for the address that is finally sent to the MMU are
respectively $a+n$ and $a+1+1\dots+1+0$.  The calculations produce
different sequences of bits to be presented to the MMU for notionally
the same address, so the two methods are incompatible and one or the
other must be preferred.  But in practice both have proved too restrictive.
It is as common, for example, for real code to step a pointer down
through an array as to step up through it, and the transformation the
compiler needs to do is prohibitively expensive.

We have accepted as engineering compromise that array access
is not for general purposes going to be constant time in this environment.
For arrays of
size $N$ the compiler can provide access in $\log N$ time in a
simple manner that is safe against all programmed
methods of calculating an index or pointer.
On our own encrypted computing platform, it is even preferable that
array access be linear time, because in order to obscure which array
element is accessed, the code must step through them
all, summing each entry in turn into an accumulator multiplied by
either 1 or 0 (encrypted) according as the entry is the one targeted or
not.  Since the multipliers are encrypted, an observer not privy to the
encryption cannot tell which multiplier is a 1 and which is a 0.  

Linear complexity code will be presented first.
To read (local) array element $\bf a[n]$ the code
tests $\bf n$ against each of $0,\dots,N{-}1$ in turn and uses a
compiler-generated address for the lookup:
\begin{center}
\begin{minipage}{0.5\textwidth}
\begin{code}
(n == 0)?a[0]:
(n == 1)?a[1]:
\dots
\end{code}
\end{minipage}
\end{center}
The equality tests are insensitive to the representation of the
same value $n$ of {\bf n}  as possibly different sequences of bits since they
take place in-processor, which discards any hidden padding and check bits.
This code always passes address $a+d$
to the MMU, where $d$ is the displacement from the base of the array and
$a$ is address ${\it sp}+k$, where $k$ is the position on the
stack allocated by the compiler for the lowest array element $\bf a[0]$.
\def\kkk{\kern-50pt}
\begin{center}
\begin{minipage}{0.5\textwidth}
\begin{code}
lw{\it r}{\it d}({\it{r}})   \,\#\rm load from address \(a+d\) with \(a\) in{\it r} to{\it r}\kkk
\end{code}
\end{minipage}
\end{center}
The
address $a$ is supplied to the base address register $r$
by a prior instruction:
\begin{center}
\begin{minipage}{0.5\textwidth}
\begin{code}
{\color{red}addi}{\it r}{\bf sp} \(k\) #{\rm add \(k\) to{\it sp} in{\it r}}
lw{\it r}{\it d}({\it{r}}) \,  #{\rm load from address \(a+d\) with \(a\) in{\it r} to{\it r}\kkk}
\end{code}
\end{minipage}
\end{center}
and this always produces the same calculation ${\it sp}+k$ for $a$.

Improving this code to $\log N$ complexity means using a binary tree
structure instead of linear lookup for the value $n$ of {\bf n},
deciding first if $n$ is below $N/2$ or above it, then on what side of
$N/4$ or $3N/4$ it is, and so on.
Code for writing to $\bf a[n]$ follows the same pattern, with store
instead of load instructions at the leaves of the binary tree or linear
code sequence.

The same code structure works for access via a pointer {\bf p}, provided
the compiler is sure which array it points into.  We have
tightened the type system of the source language (C, in our case) so
the pointer is declared along with the name of a (possibly overlarge)
array {\bf a} into which it will definitely point at runtime:
\begin{center}
\begin{minipage}{0.5\textwidth}
\begin{code}
int *p restrict a
\end{code}
\end{minipage}
\end{center}
The {\bf restrict} keyword selects the target array for the pointer.
This embellishment means a certain amount of porting has to be
done for existing code, marking out global areas into which pointers
can point.  It generally means declaring a global array from which
objects of the kind pointed to are allocated from, or declaring a
function as interior to another function where the target of the pointer
is defined as a local.  As the new pointer type is narrower than 
the original and (ideally) we make no semantic changes, 
confidence in type safety should be increased.

Then the following code does lookup via pointer {\bf p}:
\begin{center}
\begin{minipage}{0.5\textwidth}
\begin{code}
(p == a+0)?a[0]:
(p == a+1)?a[1]:
\dots
\end{code}
\end{minipage}
\end{center}
It is as insensitive to the way the pointer $\bf p$ is calculated in a
dereference $\bf*p$ as is the lookup code for $\bf a[n]$ insensitive to
the way $\bf n$ is calculated.  The code can similarly be made over to
$\log N$ complexity with a binary tree lookup structure, and converted
to write by replacing load instructions at the leaves with stores.
These constructions make pointer access the same as access via array
index.

If a size {\small\bf N} of the array is declared dynamically in the local
function, then more dynamic code must be generated. What matters is
that the compiler always causes the same calculation to be used for the
address finally sent to the MMU. The following generated code writes
reliably through pointer {\bf p}:
\begin{center}
\begin{minipage}{0.5\textwidth}
\begin{code}
for (int d=0; d<N; d++)
    if (p == a+d) \{ a[d]=x; break; \}
\end{code}
\end{minipage}
\end{center}
The address passed to the MMU is $sp+k+(0+1+\dots+1)+0$, where $a={\it
sp}+k$ is the address of array {\bf a}, and
the 1s are repeated $d$ times to address a $d$th element of the array.
The same form {\em must} be used for indexed write:
\begin{center}
\begin{minipage}{0.5\textwidth}
\begin{code}
for (int d=0; d<N; d++)
    if (n == d) \{ a[d]=x; break; \}
\end{code}
\end{minipage}
\end{center}
The calculation for the final address passed to the MMU is the same
and always accesses the same memory location as per Condition~2 of
Section~\ref{s:Intro}.

\subsection{Multi-word Types}
\label{ss:MW}

Records with named fields (`struct' in C) are treated by the compiler
as arrays and the field name is translated to an array displacement.
The declaration
\begin{center}
\begin{minipage}{0.5\textwidth}
\begin{code}
struct \{ int a; int b; \} x
\end{code}
\end{minipage}
\end{center}
declares {\bf x} with
two named fields, {\bf a} and {\bf b}, each one word wide. It occupies
two words on the stack at displacements $k$ and $k'$ (the value
$k+1$)
respectively from the stack pointer. The compiler generates accesses to
{\bf x.a} and {\bf x.b} just as it would for any local variables
situated there, by calling
\begin{center}
\begin{minipage}{0.5\textwidth}
\begin{code}
lw{\it r}{\it k}({\bf{sp}})  \#\rm load from {\bf x.a} 
\end{code}
\end{minipage}
\end{center}
to read from {\bf x.a}, for example. The address passed to the MMU is
${\it sp}+k$. To access {\bf x.b}, the address passed is ${\it
sp}+k'$ instead. With the code the compiler generates for array access,
source code attempting unsafely to access the fields of the struct as
though it were an array also works, which helps in porting.

Long atomic types such as {\bf double} are also treated this way (i.e.,
as arrays) when
distinct machine code instructions are required to access each component
words on the stack. But most platforms have double-word
load and store instructions that will fetch/write two words at once:
\begin{center}
\begin{minipage}{0.5\textwidth}
\begin{code}
ld{\it r}{\it k}({\bf{sp}})  \#\rm double word load 
\end{code}
\end{minipage}
\end{center}
and only the address of the first word is given in this case.  Registers
are indexed as pairs for this instruction, and the pair to $r$ is loaded
up by the instruction too.  
But then source code that does try to access the second word as
though the double were an array of length 2 will not work. The address
used by the MMU will be one more than the bit sequence for ${\it sp}+k$
but the address passed to the MMU by the array-oriented code generated by
the compiler will be the bit sequence for ${\it sp}+k'$, where $k'$ is
$k+1$.  Those are different.  Overall, it is safer that
the compiler {\em not} generate double word instructions.

\subsection{Short Types}

The difficulty remarked above in accessing the second word
of a {\bf double} in memory translates to a difficulty in accessing the
individual bytes of a word.  The platform usually provides instructions
that access memory a byte at a time, but the address that the MMU knows
for the second byte of a word will be one more than $a$, the sequence of
bits that the processor produces to represent the address of the word.
If the processor produces instead a sequence of bits representing $a+1$,
that in all likelihood will not be the same.  Padding
and check bits may differ.  To the processor it is all the same, but the
MMU is not privy to the secret of what bits to take notice of and what
to discard, and to it the addresses look different, and access different
bytes in memory.

For index-oriented access to the characters of a string {\bf a},
the compiler generates code that splits the character index {\bf i}
into index $\bf d$ for a word consisting of a sequence of 4 characters,
and offset $\bf j$ for the wanted character within the word:
\begin{center}
\begin{minipage}{0.5\textwidth}
\begin{code}
d = i/4; 
j = i%4;
\end{code}
\end{minipage}
\end{center}
Then the character  is obtained via an array-of-words lookup and 
arithmetic:
\begin{center}
\begin{minipage}{0.5\textwidth}
\begin{code}
(a[d] / 256\(\sp{\tt{j}}\)) % 256 \#  \(j\)th char of \(d\)th word
\end{code}
\end{minipage}
\end{center}
If the string starts in the $k$th stack position and $k'=k+d$,
the {\bf a[d]} part will result in the form of load instruction
already noted in Section~\ref{ss:MW}:
\begin{center}
\begin{minipage}{0.5\textwidth}
\begin{code}
lw \(r\) \(k'\)({\bf{sp}})  \#\rm read at offset \(k'\) from{\it sp}
\end{code}
\end{minipage}
\end{center}
In our own compiler for our own (encrypted computing) platform, we have
preferred to avoid the complication and pack characters only one to a
word, at the cost of an inefficient use of memory.  Then no special
treatment is required.  That is also helpful in the encrypted computing
environment because characters have reduced entropy (the choice is from
256 alternatives) and it is better from the point of view of defense
against stochastically-based attacks to bury them in whole words beneath
the encryption, where the rest of the word is random.

\section{Encrypted Computing}
\label{s:EC}

Although hardware aliasing takes place ubiquitously from the point of a
program running in an encrypted computing environment, that is not the
only hurdle that a compiler for that context has to overcome. In
particular, for the security of the encryption, the compiler has
to vary the underlying code and data in order to
swamp out human programming influences that could lead to statistically
based attacks.  If zero is the most common data item flowing through a
processor running human-generated code, it could pay to mount a
`plaintext' attack \cite{Biryukov2011} on the encryption supposing that
a given encrypted datum is zero.

\subsection{Address Displacement Constants}

Instead of generating a load instruction to read from a variable at
position $n$ on the stack like this (Section \ref{ss:Var}):
\begin{center}
\begin{minipage}{0.5\textwidth}
\begin{code}
lw{\it r}{\it n}({\bf{sp}})  \#\rm load from offset \(n\) from{\em sp}
\end{code}
\end{minipage}
\end{center}
the compiler will issue the instruction with a different {\em
displacement constant} $\Delta$:
\begin{center}
\begin{minipage}{0.5\textwidth}
\begin{code}
lw{\it r} {\color{red}\(\Delta\)}({\it{s}})  \#\rm load from offset \(n\) from{\em sp}
\end{code}
\end{minipage}
\end{center}
Here $\Delta$ is a previously chosen random number and the register $s$
has been pre-set with the value ${\it sp}+n-\Delta$ to accommodate this,
where {\it sp} is the nominal value for the stack pointer. The bit sequence 
passed to the MMU is from
\[
{\it sp} + n - \Delta + \Delta
\]
which\,is\,always\,the\,same\,when\,that\,calculation\,is\,repeated\,(Condition\,2,
Section\,\ref{s:Intro}).

The compiler always emits the same instruction pattern, but
it has to ensure separately that the $\Delta$ used is always the
same for the same $n$.
It maintains a vector ${\bf\Delta}$ indexed by stack location $n$,
as well as a similar vector ${\bf\Delta}_Z$ for the heap. Then
the $\Delta$ in the load instruction above is ${\bf\Delta}n$:
\begin{center}
\begin{minipage}{0.5\textwidth}
\begin{code}
lw{\it r} {\color{red}\({\bf\Delta}{n}\)}({\it{s}}) #{\rm load from offset \(n\) from{\it sp}}
\end{code}
\end{minipage}
\end{center}
The change from $n$ to ${\bf\Delta}n$ for the displacement
constant is a mark of the passage from the hardware
aliasing context to the aliasing and encrypted computing
context.  In the latter context the generated
code must vary over recompilations as further explained below.  The
change from $n$ to a randomly chosen ${\bf\Delta}n$ as an embedded
constant in the instruction is part of that.

\subsection{Content Deltas}
\label{ss:CD}

As remarked above, the stack pointer {\bf sp} does not contain the
value {\it sp} that it notionally should have but instead
is offset from that by a random value $\delta$.  That is true of the
content of every register at every point in the generated
code.  The compiler maintains a vector ${\bf\delta}_R$ of the offset
delta for content in each register $r$, varying it as it passes through
the code.  Let a non-side-effecting expression $e$ of the source
language be translated by the compiler $C_r[-]$ to machine code {\it mc} 
that targets the result for register $r$.
Let the state of the runtime machine before {\it mc}
runs be $s_0$, let the nominal value for the expression be $[e]^{s_0}$, then
\begin{align*}
({\it mc},{\bf\delta}_R) &= C_r[e] \\
s_0 &\stackrel{\it mc}{\rightsquigarrow} s_1
~~{\rm where}~ s_1(r) = [e]^{s_0} + {\bf\delta}_R\,r
\end{align*}
The `nominal value' $[e]^{s_0}$ of $e$ is 
is formalisable via a canonical construction: map a
variable $x$ in the expression to its register location $r_x$ (the
runtime value is offset by a delta ${\bf\delta}_R {r_x}$), checking the
content of $r_x$ in the state and discounting the delta to get
$[x]^{s_0}=s_0(r_x)-{\bf\delta}_R\,{r_x}$.  Arithmetic 
in the expression is formalised recursively, with
$[e_1+e_2]^{s_0}=[e_1]^{s_0}+[e_2]^{s_0}$, etc.


The compiler also maintains a vector ${\bf\delta}$ for
the delta offsets of stack contents indexed by stack location $n$, and
a vector ${\bf\delta}_Z$ for the delta offset of heap contents indexed
by heap location $n$. A window onto code generated for access to
the $n$th location on the stack shows:
\begin{center}
\begin{minipage}{0.5\textwidth}
\begin{code}
addi r{\bf sp} \(k\) \#{\rm where \(k=n-{\bf\delta}\sb{R}\,{\bf{sp}}-{\bf\Delta}{n}\)}
lw{\it r} \({\bf\Delta}{n}\)({\it{r}})  \#{\rm read \(n\)th location on stack}
\end{code}
\end{minipage}
\end{center}
The address is placed in $r$ for the load and then overwritten
with the content loaded from the location.
The MMU receives the bit sequence for the calculation
\[
{\it sp}' + k + {\bf\Delta}n
~\text{where}~ k = 
n-{\bf\delta}\sb{R}\,{\bf{sp}}-{\bf\Delta}{n}
\]
The stack pointer register {\bf sp} contains the value ${\it sp}'$
which is offset by ${\bf\delta}_R\,{\bf sp}$ from the nominal value
{\it sp} of the stack pointer.  Summing, the address has the value
${\it sp}+n$. The calculation is the same every time so the bit sequence
passed to the MMU for the address is the same every time, by Condition~2
of Section~\ref{s:Intro}. A write replaces the load by store and passes
the same bit sequence to the MMU.

The $\bf\delta$ and $\bf\Delta$ values are changed randomly by the compiler at
(just before) every point in the code where a write occurs. 
The delta for any register $r$ is
changed at every write to it.  A theorem 
in work of ours presently under review \cite{cryptoeprint:2019:084}
measures the variability in
a runtime trace in terms of its entropy viewed as a stochastic random variable
over recompilations of the same code
(the entropy is the expectation $E[-\log_2 p(T)]$ for traces $T$ that
occur with
probability $p(T)$, and it expresses the number of freely variable bits
that parametrise the trace):
\begin{theorem}
The entropy in a trace over recompilations is the sum of the entropies
of every instruction that writes that appears in it, counted once
each.
\end{theorem}
An instruction has entropy inasmuch as its effect can be varied by the
compiler from recompilation to recompilation by changing the constants
embedded in it.

Machine code instructions like load and store that merely copy from one
place to another do not contribute entropy to the trace except as they
may write to different places.
Variations in the displacement constant in the load instruction
contribute to that by changing the calculation and hence the bit
sequence passed to the MMU for the same address.

The compiler varies the contribution ${\bf\Delta} n$ in
the {\bf addi} instruction in the load sequence above,
supplying a potential 32 bits of
entropy (or 64 bits, etc., depending on the word size).  But
${\bf\Delta} n$ is fixed through reads, and only varied at
writes, so  it is old news when it runs and there is no entropy contribution 
(unless read is not preceded by write --
perhaps it is pre-loaded read-only program data).

Consideration of detail like that apart, the compiler's job in the
encrypted computing context is to do everything it can to
maximise entropy in the trace:
\begin{theorem}
The trace entropy is maximised when the compiler varies every
instruction that writes individually to the maximal extent possible
(i.e., randomly, with flat distribution) from recompilation to
recompilation.
\end{theorem}
That varies the data in the trace at runtime between different
recompilations and provides a stochastic setting in which an
attacker cannot be sure what the numerical value of the unencrypted
data should be, even in terms of a statistical tendency, because all
inputs by the programmer are swamped by the contribution from the
compiler.  The latter provides an extra, additive, uniformly distributed
input at each instruction at which it is able to.
That distribution has maximal entropy over the 32 bits (or 64 bits,
etc.) available.  The combined signal from programmer
and compiler cannot have less entropy (`Shannon's Law'
\cite{shannon48}), so it too has maximal entropy at that point, which
means that the values at that point in the trace beneath the encryption
are uniformly and evenly distributed as a probability distribution.
That means data beneath the encryption in the trace is no more
vulnerable to statistical plaintext attack than random (encrypted) data
is, though the program is designed by a human and executed
deterministically.

Structural limits on the compiler-induced variation are explained below.

\subsection{Loops}

Loops ({\bf while}, {\bf for}, backward {\bf goto}, etc.) induce 
losses in the freedom of choice for compiled code.
Let the statement compiler $C[-]$ produce code ${\it mc}$ from
statement $s$ of the source language, changing the combined database
$D=(\Delta,\Delta_Z,\delta,\delta_Z,\delta_R)$ to $D^s$ in the process. For
legibility, pairs $(D,x)$ will be written $D:x$ here:
\begin{align*}
D^s: {\it mc} &= C[D: s]
\end{align*}
The notation makes explicit that the compiler is side-effecting on $D$.

Compiling a loop {{\bf while} $e$ $s$} standardly means emitting code
{\it mc} constructed from the code ${\it mc}_e$ for $e$ and the code
${\it mc}_s$ for $s$ with the following shape:
\begin{center}
\begin{minipage}{0.5\textwidth}
\begin{code}
{\rm{start}}: {\it{mc}}\(\sb{e}\)       #{\rm compute \(e\) in \(r\)}
{\bf{beqz}} \(r\) {\rm{end}}      #{\rm goto to end if \(r\) zero}
{\it{mc}}\(\sb{s}\)             #{\rm compute \(s\)}
{\bf{b}} {\rm{start}}          #{\rm goto start}
{\rm{end}}:
\end{code}
\end{minipage}
\end{center}
The compiler produces ${\it mc}_e$, then ${\it mc}_s$, in that order:
\begin{align*}
D^e : {\it mc}_e  &= C_r\kern-1pt[D\,\,: e]\\[-0.2ex]
D^s: {\it mc}_s  &= C\,\,[D^e: s]
\end{align*}
%
%
That does not work as-is, because the code does not at the end of the loop
reestablish the deltas that existed at loop start, so the second time through,
much goes wrong.
Extra code is needed after ${\it mc}_s$, so-called `trailer' instructions.

A trailer instruction that restores the content of register $r$ to its
initial delta ${{\bf{\delta}}\sb{R}}\,r$ off the nominal value from the
final delta ${{\bf{\delta}}\sb{R}\sp{s}}r$ off nominal\,is
\begin{center}
\begin{minipage}{0.5\textwidth}
\begin{code}
addi \(r\) \(r\) \(k\)      #{\rm add \(k={{\bf{\delta}}\sb{R}}\,r-{{\bf{\delta}}\sb{R}\sp{s}}\,r\)}
\end{code}
\end{minipage}
\end{center}
Trailer instructions that restore the $n$th stack location offset also
restore the displacement constant used in load and store.  The temporary
register {\bf t0} is loaded using the final loop displacement
${\bf\Delta}\sp{s}n$, the offset of the content is modified, then stored
back using the original loop displacement ${\bf\Delta}n$, ready for
the next traverse:
\begin{center}
\hspace{0.15\textwidth}
\begin{minipage}{0.65\textwidth}
\begin{code}
addi{\bf t0 sp} \(j\)    #{\rm \(j=n-{\bf\delta}\sb{R}\,{\bf{sp}}-{\bf\Delta}\sp{s}n\)}
lw{\bf t0} \({\bf\Delta}\sp{s}n\)({\bf{t0}})  #{\rm load \(n\)th stack location}\vspace{0.75ex}
addi{\bf t0 t0} \(k\)    #{\rm modify by \(k={{\bf{\delta}}n-{{\bf{\delta}}\sp{s}}}n\)}\vspace{0.75ex}
addi{\bf t1 sp} \(l\,\)    #{\rm \(l=n-{\bf\delta}\sb{R}\,{\bf{sp}}-{\bf\Delta}n\)}
sw \({\bf\Delta}n\)({\bf{t1}}){\bf t0}  \,\,#{\rm store \(n\)th stack location}
\end{code}
\end{minipage}
\end{center}
The first two instructions are exactly the read stack code of
Section~\ref{ss:CD}, and the last two instructions are the corresponding
write stack code.  The stack pointer does not change from beginning to
end of the loop, nor does the offset $\delta\sb{R}{\bf{sp}}$ of its
content from the nominal value {\it sp}, so it appears unaltered
through that code.

The instructions above are determined by choices of deltas by the compiler
for earlier instructions.  It is impossible to execute the trailer
instructions without traversing the loop body, which will execute those
earlier instructions, so these trailer instructions are always `old
news' and introduce no entropy into the trace.

\subsection{Conditionals}

The synchronisation problem described above occurs
wherever two distinct control paths join, and after an
if-then-else block in particular.  The final deltas in the two
branches/paths must be
equalised.  The compiler emits final `trailer' instructions 
for that purpose. Code {\it mc} for {\bf if}\,$e$\,$s_1$\,{\bf else}\,$s_2$
standardly has shape:
\begin{center}
\begin{minipage}{0.5\textwidth}
\begin{code}
{\rm{start}}: {\it{mc}}\(\sb{e}\)       #{\rm compute \(e\) in \(r\)}
{\bf{beqz}} \(r\) {\rm{else}}  \,\,   #{\rm goto to else if \(r\) zero}
{\it{mc}}\(\sb{s\sb1}\)            #{\rm compute \(s\sb1\)}
{\bf{b}} {\rm{end}}           #{\rm goto end}
{\rm{else}}: {\it{mc}}\(\sb{s\sb2}\)       #{\rm compute \(s\sb2\)}
{\rm{end}}:
\end{code}
\end{minipage}
\end{center}
where the order the compiler works in is ${\it mc}_e$, ${\it mc}_{s_1}$,
${\it mc}_{s_2}$:
\begin{align*}
D^e \,\,: {\it mc}_e  \,\,&= C_r\kern-1pt[D\,\,\,\,: e\,]\\[-0.2ex]
D^{s_1}: {\it mc}_{s_1}  &= C\,\,[D^e\,\,: s_1]\\[-0.2ex]
D^{s_2}: {\it mc}_{s_2}  &= C\,\,[D^{s_1}: s_2]
\end{align*}
But that code does not work because it also needs trailer instructions after
${\it mc}_{s_2}$ to set the final deltas in the `else' branch equal to
the final deltas in the `then' branch. Each has the same pattern as
a trailer instruction for loops.

Other notable places where trailer instructions are required are the
label targets of {\bf goto}s, and at {\bf return} from functions. Calls
of interior functions also require `trailers', but before the call, because
the context in which the function was defined must be reestablished
(effectively a `come from' imperative).

\section{Implementation}
\label{s:Imp}

\def\ME[#1]{#1}
\def\la{\(\leftarrow\)}
\begin{table}[!tbp]
\caption{Trace for Ackermann(3,1)}
\label{tab:3}
\flushleft
\begin{tabular}{@{}l@{}r@{}}
\begin{minipage}{0.6\textwidth}
\begin{alltt}\scriptsize
{\rm{PC}}  {\rm{instruction}}                     {\rm{trace updates}}
\dots
35  addi t0 a0      \ME[-86921031]       t0 \la \ME[-86921028]
36  addi t1 zer     \ME[-327157853]      t1 \la \ME[-327157853]
37  beq  t0 t1  2   \ME[240236822]                  
38  addi t0 zer     \ME[-1242455113]     t0 \la \ME[-1242455113]
39  b 1                                             
41  addi t1 zer     \ME[-1902505258]     t1 \la \ME[-1902505258]
42  xor  t0 t0  t1  \ME[-1734761313] \ME[1242455113] \ME[1902505258]
                                    t0 = \ME[-17347613130]
43  beqz t0 9       \ME[-1734761313]                 
53  addi sp sp      \ME[800875856]       sp \la \ME[1687471183] 
54  addi t0 a1      \ME[-915514235]      t0 \la \ME[-915514234] 
55  addi t1 zer     \ME[-1175411995]     t1 \la \ME[-1175411995]
56  beq  t0 t1  2   \ME[259897760]                   
57  addi t0 zer     \ME[11161509]        t0 \la \ME[11161509]   
\dots
143 addi v0 t0      \ME[42611675]       \fbox{v0 \la \ME[13]} #{\rm result}
\dots
147 jr  ra
STOP
\end{alltt}
\end{minipage}
&
\begin{minipage}[t]{0.38\textwidth}
\scriptsize%
\begin{tabular}{@{}|l@{~}l@{~}l|@{}}
\hline
\multicolumn{3}{@{}|c|@{}}{\vbox to 3ex{}Legend}\\
\hline
\em op.& {\em fields} & {\em semantics}\\
\hline
\vbox to 3ex{}addi &$\rm r_0$\,$\rm r_1$\,$k$ 
        &$r_0 \leftarrow r_1+k$\\
b   &$i$
        &${\it pc}\leftarrow {\it pc}+i$\\
beq& $\rm r_1$\,$\rm r_2$\,$i$
        &${\rm if}\,r_1 {=} r_2\,{\rm then}\,{\it pc}{\leftarrow}{\it pc}{+}i$\\
jr   &$\rm r$
       &${\it pc} \leftarrow r$\\
xor &$\rm r_0$\kern1pt$\rm r_1$\kern1pt$\rm r_2$
        &$r_0\leftarrow (\kern-1ptr_1{+}k_1)\mathop{\widehat{ }}(\kern-1ptr_2{+}k_2){-}k_0$\\
                & \kern1pt$k_1$\kern1pt$k_2$\kern1pt$k_0$&\\
\hline
\multicolumn{2}{@{}|l}{\em register} & {\em semantics}\\
\hline
\multicolumn{2}{@{}|l}{a0,a1,\dots} & function argument\\
\multicolumn{2}{@{}|l}{pc} & program counter\\
\multicolumn{2}{@{}|l}{ra} & return address\\
\multicolumn{2}{@{}|l}{sp} & stack pointer\\
\multicolumn{2}{@{}|l}{t0,t1,\dots} & temporaries\\
\multicolumn{2}{@{}|l}{v0,v1,\dots} & return value\\
\multicolumn{2}{@{}|l}{zer}& null placeholder\\
\hline
\multicolumn{2}{@{}|l}{\em lexicon} & {\em semantics}\\
\hline
\multicolumn{2}{@{}|l}{$i$} & program count increment\\
\multicolumn{2}{@{}|l}{$k$} & instruction constant\\
\multicolumn{2}{@{}|l}{$r$} & content of r\\
\hline
\end{tabular}
\end{minipage}
\end{tabular}
\end{table}
 
Our own prototype C compiler \url{http://sf.net/p/obfusc} 
covers {\sc ansi} C and GNU C extensions, including
statements-as-expressions and expressions-as-statements, gotos, arrays,
pointers, structs, unions, floating point, double integer and floating
point data.  It is missing {\bf longjmp} and efficient strings
({\bf char} and {\bf short} are the same as {\bf
int}), and global data shared across code units (a linker issue due
to the who-decides-the-delta conundrum for code compiled first but
referencing as-yet uncompiled external global data).
%

A trace\footnote{Initial and final content offset deltas
are set to zero here, for readability.} of the Ackermann
function\footnote{Ackermann
C code: {\bf int} A({\bf int} m,{\bf int} n) \{ {\bf if} (m == 0) {\bf
return} n+1; {\bf if} (n == 0) {\bf return} A(m-1, 1); {\bf return}
A(m-1, A(m, n-1)); \}.} \cite{Sundblad71} compiled by that compiler
is shown in Table~\ref{tab:3}.  The trace illustrates how the compiler's
variation of the delta offsets for register content through the code results
in randomly generated constants embedded in the instructions and
randomly offset runtime data.

\begin{table}[!tp]
\caption{Trace for sieve showing hidden bits in data (right).
Stack read instruction lines are in red, address base for lookup and
address displacement in blue.}
\label{tab:4}
\flushleft
\begin{tabular}{@{}l@{}r@{}}
\begin{minipage}{0.95\textwidth}
\begin{alltt}\scriptsize
{\rm{PC}}     {\rm{instruction}}                      {\rm{trace updates}} | {\rm{hidden}}
\dots
22340   addi t1  sp  \ME[-418452205]       t1 \la \ME[-877254954|1532548040]
22360   bne  t0  t1  84
{\color{red}22384   addi t1  sp  \ME[-407791003]       t1 \la \ME[{{\color{blue}-866593752|1532548040}}]} #{\rm read local array}
{\color{red}22404   lw   t0  \ME[{{\color{blue}866593746}}](t1)        t0 \la \ME[-866593745|1800719299]} # a[7]{\rm at}{\it sp}{\rm+40}
22424   addi t0  t0  \ME[-1668656853]      t0 \la \ME[1759716698|1081155516]
22444   b    540
22988   addi t1  zer \ME[1759716697]       t1 \la \ME[1759716697|1325372150]
23008   bne  t0  t1  44
\dots
{\color{red}23128   addi t0  sp  \ME[-1763599776]      t0 \la \ME[{{\color{blue}2072564771|-1935092797}}]} #{\rm read local variable}
{\color{red}23148   lw   t0  \ME[{{\color{blue}-2072564772}}](t0)      t0 \la \ME[2072564779|-1773201679]} # i{\rm at}{\it sp}{\rm+45}
23168   addi t0  t0  \ME[1723411350]       t0 \la \ME[-498991167|-981581771]
23188   addi t0  t0  \ME[-1862832992]      t0 \la \ME[1933143137|-1629507929]
23208   addi v0  t0  \ME[-1933143130]     \fbox{v0 \la \ME[7]}        \:|1680883739 # return
\dots
23272   jr   ra
STOP
\end{alltt}
\end{minipage}
&
\end{tabular}
\end{table}

Running a Sieve of Eratosthenes program\footnote{Sieve C code:
{\bf int}
S({\bf int} n\,) \{ {\bf int} a[N]=\{[0\dots N-1]=1,\};
    {\bf if} (n$>$N$||$n$<$3) {\bf return} 0;
    {\bf for} ({\bf int} i=2; i$<$n; ++i) \{
        {\bf if} (!a[i]) {\bf continue};
        {\bf for} ({\bf int} j= 2*i; j$<$n; ++j) a[j]=0;
        \}; 
    {\bf for} ({\bf int} i=n-1; i$>$2; {-}{-}i)
        if (a[i]) {\bf return} i;
    {\bf return} 0;
\}
.
}
for primes shows how memory access is affected by address displacement
constants.  The final part of the trace  is shown in Table~\ref{tab:4} with
two stack reads in red and the address base and address displacement in blue.
The assignments to these stack locations are up-trace and do have the
same address base and displacements as in the later reads:
\begin{code}
19300   addi t1  sp  \ME[-407791003]    t1 \la \ME[{\color{blue}-866593752}|{\color{blue}1532548040}]
19320   sw   \ME[{{\color{blue}866593746}}](t1) t0      mem[\ME[-6|-712377144]]
                                      \la \ME[-866593745|1800719299]
\dots
20884   addi t1  sp  \ME[-1763599776]   t1 \la \ME[{\color{blue}2072564771}|{\color{blue}-1935092797}]
20904   sw   \ME[{{\color{blue}-2072564772}}](t1) t0    mem[\ME[-1|1518992593]]
                                      \la \ME[2072564779|-1773201679]
\end{code}
The memory addresses -6, -1 reflect that the stack grows down from 
top of memory (-1), but it is the combinations -6$|$-712377144 and -1$|$1518992593
of address and hidden bits together that are looked up by the MMU.

\section{Conclusion}

This paper has described the compilation of imperative low level source code
for a platform that has hardware aliasing with hidden
determinism.  The technique depends on the compiler controlling
exactly the address displacement and address base for load and store
instructions so that they are always the same for repeat
accesses to what is intended to be the same memory location.  That means
they are either copies or the calculations for them exactly reprise the
earlier calculations.  It is surprising that the trick can be worked in
such a systematic way, but it can further be extended to both generate
and cover for displacements in the data content of registers and memory
in the context of encrypted computing.  There it is essential for the
security of encrypted data passing through the machine that it may
have been varied by a maximal entropy input from the compiler, which
swamps statistical biases that arise from human programming.

The mechanism described in this paper is log or linear complexity for array
access.  Constant complexity would be possible but at the price of
making index- and pointer-based accesses mutually incompatible, which
would necessitate invasive changes in ported source code.

In the encrypted computing context, the existence of this kind of
compiler validates those processor designs that pass encrypted addresses
as well as data to the memory unit, which remains fully ignorant of the
encryption.

\begin{small}
\bibliography{rst2019}

\begin{thebibliography}{10}

\bibitem{Barr98}
Barr, M.:
\newblock Ch.\ 6, {M}emory.
\newblock In Oram, A., ed.: Programming Embedded Systems in C and C++.
\newblock 1st edn. O'Reilly \& Associates, Inc., Sebastopol, {CA} (1998)
  64--92

\bibitem{Glosserman85}
Glosserman, P.:
\newblock {Quarterdeck} Expanded Memory Manager: {QEMM}, instant power for 386,
  486 or {Pentium PCs}.
\newblock Quarterdeck Office Systems (1985)

\bibitem{ostrovsky1990}
Ostrovsky, R.:
\newblock Efficient computation on oblivious {RAM}s.
\newblock In: Proc.\ 22nd Ann.\ {ACM} Symp.\ Th.\ Comp. (1990)  514--523

\bibitem{ostrovsky1996}
Goldreich, O., Ostrovsky, R.:
\newblock Software protection and simulation on oblivious rams.
\newblock Journal of the {ACM} ({JACM}) \textbf{43}(3) (1996)  431--473

\bibitem{ostrovsky2013}
Lu, S., Ostrovsky, R.:
\newblock Distributed oblivious {RAM} for secure two-party computation.
\newblock In Sahai, A., ed.: Proc.\ 10th Th.\ Cryptog.\ Conf.\ ({TCC}'13).
  Volume 7785 of LNCS.
\newblock Springer, Berlin/Heidelberg (2013)  377--396

\bibitem{Gruhn2013}
Gruhn, M., M\"uller, T.:
\newblock On the practicability of cold boot attacks.
\newblock In: 8th Int.\ Conf.\ Availability, Reliability, Sec.\ ({ARES}'13).
  (2013)  390--397

\bibitem{fletcher2012}
Fletcher, C.W., van Dijk, M., Devadas, S.:
\newblock A secure processor architecture for encrypted computation on
  untrusted programs.
\newblock In: Proc.\ 7th ACM Scalable Trusted Comp.\ Workshop ({STC}'12), New
  York, {ACM} (2012)  3--8

\bibitem{BB13a}
Breuer, P.T., Bowen, J.P.:
\newblock A fully homomorphic crypto-processor design: Correctness of a secret
  computer.
\newblock In J{\"u}rjens, J., Livshits, B., Scandariato, R., eds.: Proc.\ 5th
  Int.\ Symp.\ Eng.\ Sec.\ Soft.\ Sys.\ ({ESSoS}'13). Number 7781 in LNCS,
  Berlin/Heidelberg, Springer (2013)  123--138

\bibitem{oic}
Tsoutsos, N., Maniatakos, M.:
\newblock Investigating the application of one instruction set computing for
  encrypted data computation.
\newblock In Gierlichs, B., Guilley, S., Mukhopadhyay, D., eds.: Proc.\ Sec.,
  Priv.\ Appl.\ Cryptog.\ Eng.\ ({SPACE}'13).
\newblock Springer, Berlin/Heidelberg (2013)  21--37

\bibitem{heroic}
Tsoutsos, N., Maniatakos, M.:
\newblock The {HEROIC} framework: Encrypted computation without shared keys.
\newblock {IEEE} Trans.\ {CAD} Integ.\ Circ.\ Sys. \textbf{34}(6) (2015)
  875--888

\bibitem{BB16a}
Breuer, P.T., Bowen, J.P.:
\newblock A fully encrypted microprocessor: The secret computer is nearly here.
\newblock Procedia Comp. Sci. \textbf{83} (2016)  1282--1287

\bibitem{BB16b}
Breuer, P.T., Bowen, J.P., Palomar, E., Liu, Z.:
\newblock A practical encrypted microprocessor.
\newblock In Callegari, C.,  et~al., eds.: Proc.\ 13th Int.\ Conf.\ Sec.\
  Cryptog. ({SECRYPT}'16). Volume~4., Portugal, {SCITEPRESS} (2016)  239--250

\bibitem{DBLP:journals/corr/abs-1811-12365}
Breuer, P.T., Bowen, J.P.:
\newblock (un)encrypted computing and indistinguishability obfuscation.
\newblock CoRR \textbf{abs/1811.12365} (2018) (Extended abstract) Princip.\
  Sec.\ Compil.\ Track ({PRiSC}'19), 46th {ACM} {SIGPLAN} Symp.\ Princip.\
  Prog.\ Lang.\ (POPL'19).

\bibitem{Goldwasser1982}
Goldwasser, S., Micali, S.:
\newblock Probabilistic encryption \& how to play mental poker keeping secret
  all partial information.
\newblock In: Proc.\,14th Ann.\ {ACM} Symp.\,Th.\ Comp. {STOC}'82, {ACM} (1982)
   365--377

\bibitem{BB18c}
Breuer, P.T., Bowen, J.P., Palomar, E., Liu, Z.:
\newblock On security in encrypted computing.
\newblock In Naccache, D.,  et~al., eds.: Proc.\ 20th Int.\ Conf.\ Info.\
  Comms.\ Sec.\ {(ICICS'19)}. Number 11149 in LNCS, Cham, Springer (2018)
  192--211

\bibitem{cryptoblaze18}
Irena, F., Murphy, D., Parameswaran, S.:
\newblock Cryptoblaze: A partially homomorphic processor with multiple
  instructions and non-deterministic encryption support.
\newblock In: Proc.\ 23rd Asia S.\ Pac.\ Design\ Autom.\ Conf.\ ({ASP-DAC}),
  Los Alamitos, CA, USA, IEEE (2018)  702--708

\bibitem{BB18b}
Breuer, P.T., Bowen, J.P., Palomar, E., Liu, Z.:
\newblock Superscalar encrypted {RISC}: The measure of a secret computer.
\newblock In: Proc.\ 17th Int.\ Conf.\ Trust, Sec.\ Priv.\ Comp.\ Comms.\
  ({TrustCom}'18), CA, USA, {IEEE} Comp. Soc. (2018)  1336--1341

\bibitem{BB14c}
Breuer, P.T., Bowen, J.P.:
\newblock Avoiding hardware aliasing: Verifying {RISC} machine and assembly
  code for encrypted computing.
\newblock In: Proc.\ 2nd {IEEE} Workshop Reliability Sec.\ Data Anal.
  ({RSDA}'14), {IEEE} Int.\ Symp.\ Soft.\ Reliability Eng.\ Workshops
  ({ISSREW}'14), Los Alamitos, {CA}, {IEEE} Computer Society (2014)  365--370

\bibitem{ansi99}
{ISO/IEC}:
\newblock Programming languages -- {C}.
\newblock 9899:201x Tech.\ Report n1570, International Organization for
  Standardization (2011) {JTC 1, SC 22, WG 14}.

\bibitem{Biryukov2011}
Biryukov, A.:
\newblock Known plaintext attack.
\newblock In van Tilborg, H.C.A., Jajodia, S., eds.: Encyclopedia of
  Cryptography and Security.
\newblock Springer, Boston, MA (2011)  704--705

\bibitem{cryptoeprint:2019:084}
Breuer, P.T.:
\newblock An information obfuscation calculus for encrypted computing.
\newblock Cryptology ePrint Archive, Report 2019/084 (2019)

\bibitem{shannon48}
Shannon, C.E.:
\newblock A mathematical theory of communication.
\newblock Bell System Technical Journal \textbf{27}(3) (1948)  379--423

\bibitem{Sundblad71}
Sundblad, Y.:
\newblock The {Ackermann} function.\ a theoretical, computational, and formula
  manipulative study.
\newblock {BIT} Num. Math. \textbf{11}(1) (1971)  107--119

\end{thebibliography}
\end{small}

\end{document}